\documentstyle[psfig]{caosp}
\begin{document}
\pubyear{1998}
\volume{27}
\firstpage{261}
\title{Effective temperatures of CP stars}
\author{N.A. Sokolov}
\institute{Central Astronomical Observatory at Pulkovo,
 St.Petersburg 196140, Russia}
\date{October 29, 1997}
\maketitle
\begin{abstract}
A new method of $T_{\rm eff}$ determination for
CP stars is proposed. The method is based on the fact that the
slope of the energy distribution in the Balmer continuum near
the Balmer jump is identical for "normal" main sequence stars and
for CP~stars with the same $T_{\rm eff}$.
It is shown that the $T_{\rm eff}$ of CP stars derived
by this method are in good agreement with those derived
by other methods.
\end{abstract}

\section{Introduction}

A review of the various methods of effective temperature
determination for chemically peculiar (CP) stars shows once
more the difficulty of deriving the $T_{\rm eff}$ of these stars.
If one uses methods taking into consideration the blanketing
effect, the temperature obtained is close to the effective one.
In the infrared flux method (IRFM) first proposed by Blackwell \&
Shallis~(1977), a monochromatic flux is measured in the infrared
region to minimize the blanketing effect. The method proposed
by Stepien \& Dominiczak~(1989) takes into account the blanketing
effect as well.
The photometric methods may be useful, since it is possible to apply 
a correction to the color (or model) temperature and to give 
relatively good estimates of $T_{\rm eff}$ (Hauck \& North~1993).
Another way is to use an observational parameter which is not
affected by peculiarities and can be applicable both to
the "normal" main sequence stars and to the CP stars.
The Balmer continuum slope near the Balmer jump ($\it\Phi_{\rm u}$) 
can be a useful tool for determination of $T_{\rm eff}$ for 
CP stars (Sokolov 1995).
The determination of the effective temperatures of
CP stars using the $\it\Phi_{\rm u}$ is discussed.

\section{The method description}

It is well known that the first order
difference between energy distribution of "normal" main sequence
stars and peculiar stars is caused by extra blocking of the
flux in the far-UV and the redistribution of it in the longer
wavelengths.
The method presented here assumes the existence of a wavelength region,
between the far-UV and visual regions, where the energy distribution is the
same for ``normal" and for CP~stars.
The comparison of the observed energy distribution with
the best fitting model for twelve CP stars given by Stepien \&
Dominiczak (1989), supports this assumption and shows that 
$\it\Phi_{\rm u}$ is identical both for the observed energy distribution of 
CP stars and for the models (see Fig.~2 of Stepien~\& 
Dominiczak~1989).
From the theoretical point of view,
the computations carried out by Leckrone et al.~(1974) and by
Muthsam (1978, 1979) show that there exists a wavelength
$\lambda_{\rm tr}$ such that for $\lambda < \lambda_{\rm tr}$
the flux of a CP star is lower than that of a
normal star with the same $T_{\rm eff}$, whereas for
$\lambda >\lambda_{\rm tr}$ it is enhanced.
Note that the location of $\lambda_{\rm tr}$ is in the Balmer
continuum near the Balmer jump, which is used for the determination
of $T_{\rm eff}$ (Sokolov~1995).

Based on the theoretical prediction, as well as on the fact that
the $\it\Phi_{\rm u}$ of models and
CP stars is identical, this observational parameter is used to
determine the $T_{\rm eff}$ of CP stars.
The temperature calibration of B, A and F main sequence stars
given by Sokolov~(1995) is applied to the CP stars as well.

\section{Results and discussion}

The calibration curve described above was applied to
50 CP2 stars from the catalog of stellar spectrophotometry
(Adelman et al. 1989) and to 18 CP2 stars from the Pulkovo
spectrophotometric catalog of bright stars (Alekseeva et al. 1996).
To test the validity of the proposed method of determination of
$T_{\rm eff}$ for CP2 stars, the temperatures derived from
$\it\Phi$$_{\rm u}$ were compared with those derived from the IRFM,
from the method of Stepien~\&~Dominiczak~(1989)
and from the (B2-G) color index of Geneva photometry.

In the literature we found five papers concerning CP2 stars
for which the effective temperature is derived using the IRFM.
The values of $T_{\rm eff}$ derived from $\it\Phi$$_{\rm u}$
are compared with those obtained by IRFM for 13 common stars
(see Fig.~1a).
One can see that the agreement appears to be very good.
So, the mean effective temperature
difference is $\Delta T_{\rm eff}~=~
T_{\rm eff}$($\it\Phi$$_{\rm u}$)~-~$T_{\rm eff}$(IRFM)~
=~41$\pm$127~K, with a linear correlation coefficient {\it r}~=~0.972,
and $\alpha$~=~0.904 for the slope of the regression line of
$T_{\rm eff}$($\it\Phi$$_{\rm u}$) versus
$T_{\rm eff}$(IRFM). 
\begin{figure} [t]
\centering{
\vspace{-1.1cm}
\hspace*{-1.1cm}
\vbox{\psfig{figure=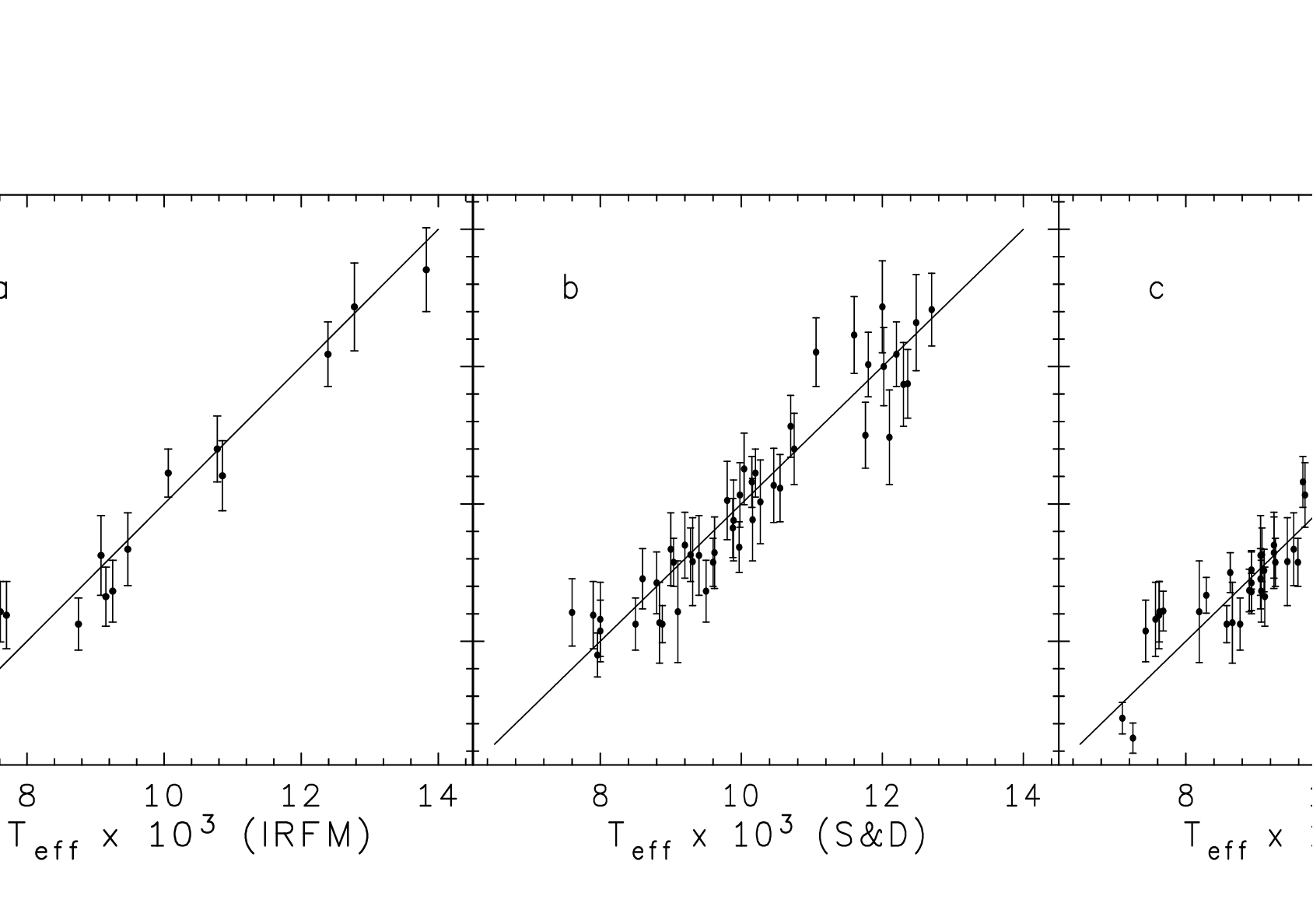,angle=270,height=6.0cm}}\par
}
{\small Fig. 1: Comparison of $T_{\rm eff}$ derived from
$\it\Phi_{\rm u}$ with those derived by infrared flux
method - (a), by the method proposed by Stepien \& Dominiczak - (b)
and from (B2-G) color index of Geneva photometry - (c).}
\end{figure}

Stepien~\&~Dominiczak (1989) proposed a new method to determine
the effective temperatures of CP2 stars.
In order to have a wider sample of stars for the comparison,
this method was applied to the model temperatures ($T_{\rm M}$)
obtained by Adelman~(1985).
The value of $T_{\rm M}$ was calculated as the average of $T$(PC) and
$T$(BJ) for all stars of Table~2 of Adelman~(1985).
After that, this mean value of $T_{\rm M}$ was corrected for
the blanketing effect by using Eq.~12 from the paper of Stepien~\&~Dominiczak. 
The resulting temperatures
($T_{\rm eff}$(S\&D)) are compared with $T_{\rm eff}$ derived from
$\it\Phi$$_{\rm u}$. Figure~1b gives a plot of 
$T_{\rm eff}$($\it\Phi$$_{\rm u}$) versus $T_{\rm eff}$(S\&D) 
for 47 common stars.
Basically, there are no systematic differences between the two sets
of data, as confirmed by the following results:
$\Delta$$T_{\rm eff}$~=~38$\pm$69~K, {\it r}~=~0.952, $\alpha$~=~0.965.

In order to estimate photometrically the effective temperatures
of CP2 stars the (${\it B}{\rm 2-}{\it G}$) color index was used.
Figure~1c gives a plot of the $T_{\rm eff}$ derived from
$\it\Phi_{\rm u}$ versus $T_{\rm eff}$(${\it B}{\rm 2-}{\it G}$)
for the 59 common stars.
One can see that there is no systematic difference between the two
sets of data, though the scatter of the points on Figure~1c
is rather high (up to 1000~K), especially for the stars with
$T_{\rm eff}>9500$~K.
For the stars in our sample we have 
$\Delta T_{\rm eff}$~=~102$\pm$76~K, {\it r}~=~0.938,
and $\alpha$~=~0.975.

Generally, there is no significant systematic difference between
the temperatures derived from $\it\Phi$$_{\rm u}$ and those
derived from fluxes by other methods. The temperature calibration
derived for B, A and F main sequence stars is applicable to CP2
stars as well. The temperature derived from $\it\Phi$$_{\rm u}$
for CP2 stars may be identified with their effective one,
because the influence of the stars's peculiarity on
the Balmer continuum slope near the Balmer jump is negligible.
In our study only CP2~stars were used, but this method can be
extended to other types of CP~stars, for which the blanketing effect
is less pronounced: for them, the temperature derived from $\it\Phi_{\rm u}$ 
should then be even closer to the effective one.

\end{document}